# Numerical study of the electrical response from a mechanical model describing interactions into a slipping multi-contact interface

E. Chevallier,[1,a]

[1] *Laboratoire « Physique des Systèmes Complexes » (P.S.C. - EA 4663), Université de Picardie Jule Vernes, 33 rue Saint Leu, 80 0039 Amiens, FRANCE.*

We report a numerical study accounting for the electrical response of a slipping multi-contact interface and of their variations with the sliding speed. We characterize these variations by analyzing the voltage noise across the contact from a mechanical model based on the roughness interaction. Our model focuses only on the mechanical aspects of the slipping interface and explores its electrical consequences (electrical noise) versus the sliding speed. Numerical results are compared with experimental data of the resistance noise versus sliding speed: we find a good agreement between theory and experiments, which opens ways to the development of the electro-mechanic devices monitoring.

Keywords: electromechanical systems, slipping interface, electrical contact, tribological engineering.

**I. INTRODUCTION**

In industrial machines, the transfer of electrical signals as a current or any monitoring signal is carried out through electromechanical devices as a sliding contact [1, 2, 3], coupling two components in relative motion. The coupling of moving components through a sliding contact is widely used in many fields where the electrical contact is established by a wire in a groove, a brush, or a skate on a metallic surface that can be lubricated. One of the main problems of these electromechanical systems is the electrical contact noise: signals sent through the contact always emerge with a noisy component more or less important. The optimization of the parameters connected with both the electrical and mechanical efficiencies of the contact appears then as a fundamental challenge. The performance of these systems depends strikingly on our understanding of the interface interactions, which are the physical processes involved in the contact. Only a deep understanding of such processes allows the identification of reliable criteria of the contact's quality. That is why we focus here on the electrical resistance properties of the interface.

In [5] Chevallier and al. have proposed a theoretical description accounting for the interface voltage noise from aging measurements of an Au/Au slipping ring/wire interface. From this work, we built a numerical model which can report on the mechanical influence on the voltage noise. Thus, we propose an approach of the electrical properties of the sliding interface versus the sliding speed. This model is based on a simple mechanical behavior where the asperities – i.e. the contact spots – are (i) mechanically and electrically independent and where (ii) the fluctuation of the contact number inside the interface during sliding is piloted by the three functionality parameters: the steady force $\vec{W}$, sliding speed V, and roughness.

---

[a] Corresponding author, eddy.chevallier@u-picardie.fr

The surface morphology of the ring is modeled as a steady geometry with a random part to mimic its roughness [6, 7, 8]. The low-scale random part of the geometry accompanies a deterministic higher scale contribution (always present), as for instance a machined surface [9]. The contact zone is treated as a multi-contact interface [10, 11], so the contact zone conductance is then proportional to the number of asperities couples. During sliding, the number of couples of contacting asperities varies randomly around a mean value $\bar{N}$ depending on the steadying force [10, 11, 12, 13]. Note we have identified several contributions to the voltage fluctuations in [5], so we just focus here on the mechanical aspect to develop our model.

**II. MECHANICAL MODEL**

As we stated earlier, the spot displacement depends of main functionality parameters: the sliding speed V, and the steady force – or contact load – $\vec{W}$. During its movement, the behavior of a lonely contact spot is like a switch function: sometimes the contact is made - switch function equal to 1 - and sometimes contact is not made - switch function equal to 0 -. When the contact is realized the switch function is equal to 1 during a contact time that we call ε, and respectively when the contact is not realized, the function is equal to 0 during a no-contact time that we call Δ. Mathematically we can express the switch function as a sum of boxcar function where each boxcar is linked to an uncoupling moment $t_i$ and a couple of the specifics duration $\varepsilon_i$ and $\Delta_i$. So, we have:

$$K_a(t) = \sum_i \theta(t - t_i - \Delta_i)(1 - \theta(t - t_i - \Delta_i - \varepsilon_i)) \qquad \text{[Eq. 01]}$$

The index "a" of our switch function here means that it is defined just for one spot. Thus, if we associate all spots contributions we describe the phenomenological aspect of the mechanical interactions into the sliding interface.

To go further, we have to determine the coupling and uncoupling instants. To determine these ones, we have to study specifically the profile curvature on which the spot moves and the mechanical conditions of its motion. Based on the mechanical assumptions schematized on figure 1 we can determine the uncoupling instants as the solution of the following equation:

$$\vec{R_N} = -(\vec{W_R} + \vec{F_i}) = 0 \qquad \text{[Eq. 02]}$$

Or

$$\frac{V_\tau^2}{R_\gamma(\vec{x}(t))} = \frac{W_R(\vec{x}(t))}{m} \qquad \text{[Eq. 03]}$$

Where $V_\tau$ is the horizontal sliding speed projected on the $\vec{\tau}$ axe, m the mass, $W_R$ the projection of $\vec{W}$ on the $\vec{R}$ axe, $R_\gamma$ the curvature radius at the x(t) position and $\vec{F_i}$ the inertia force generated by the profile curvature.

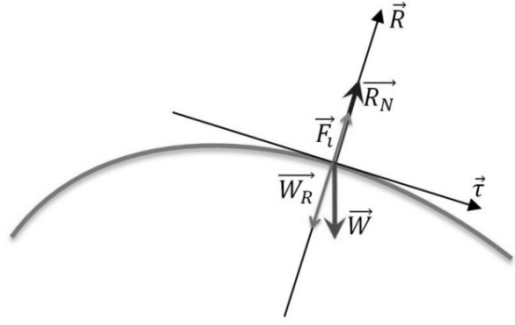

FIG. 1. Schematic visualization of the mechanical conditions of the contact spot on a profile. The contact spot is mechanically considered as a dimensionless point on which the forces are concentrated.

Once got the uncoupling instants, we have to get the coupling moments which follow. Several approaches are possible but in our case we build a simplified behavior based on free-fall, with an effective gravity parametrized by the steady force W and the mass m: $g^* = (g + W/m)$. These conditions once introduce finally give us the coupling instants which follow the uncoupling ones. Note that the mass "m" must be understood as a parameter introduced depending on the context of the application: in the case of a ring/wire interface we can consider the contact as punctual and so consider the mass "m" as the wire mass.

The behavior of a contact spot that follow the mechanical considerations established previously can be observed on figure 2 where is represented the influence of the sliding speed on the contact and no-contact durations. A first comparison of the mechanical behavior for two different sliding speeds indicates that, as we suspected it, more the sliding speed, earlier the uncoupling instant, and later the coupling instant which follows. This leads to an increase of the no-contact duration and a decrease of the contact duration.

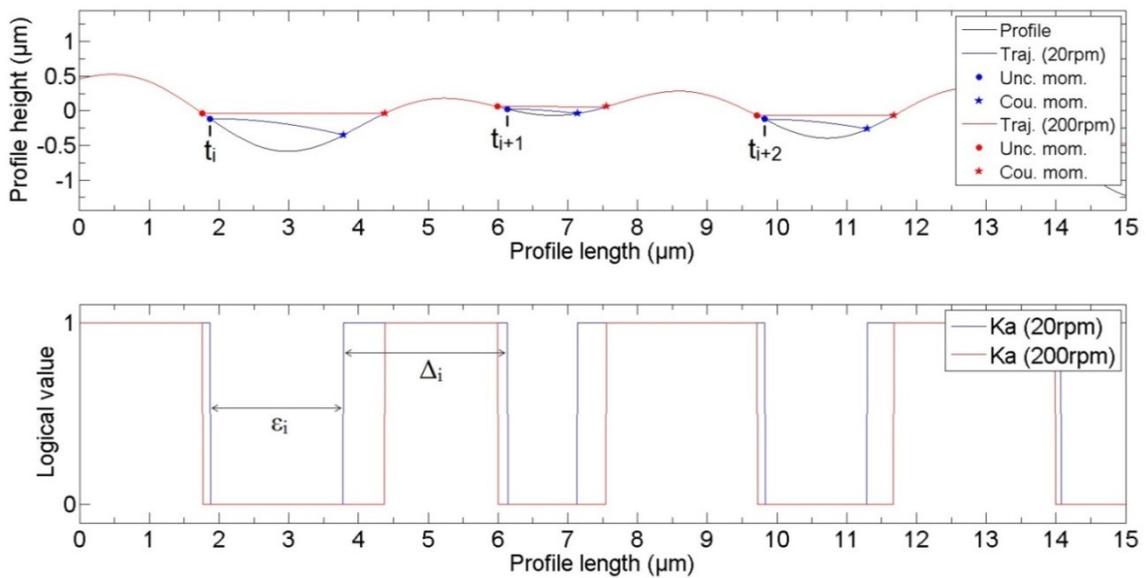

FIG. 2. Up: Spot trajectory visualization on a profile for two values of sliding speed, 20 rpm in blue color, 200 rpm in red, with their respective coupling and uncoupling moments. Down: associated switch functions of the spot trajectories.

Once each switch functions being determined, we can summarize all the spots contribution and get the associated "contact noise", which indicate the number of contact inside the interface (fig. 3) during motion. The contact number fluctuations, aka contact noise, have a mean value $\bar{N}$ and a standard deviation $\sigma_N$ for a given sliding speed.

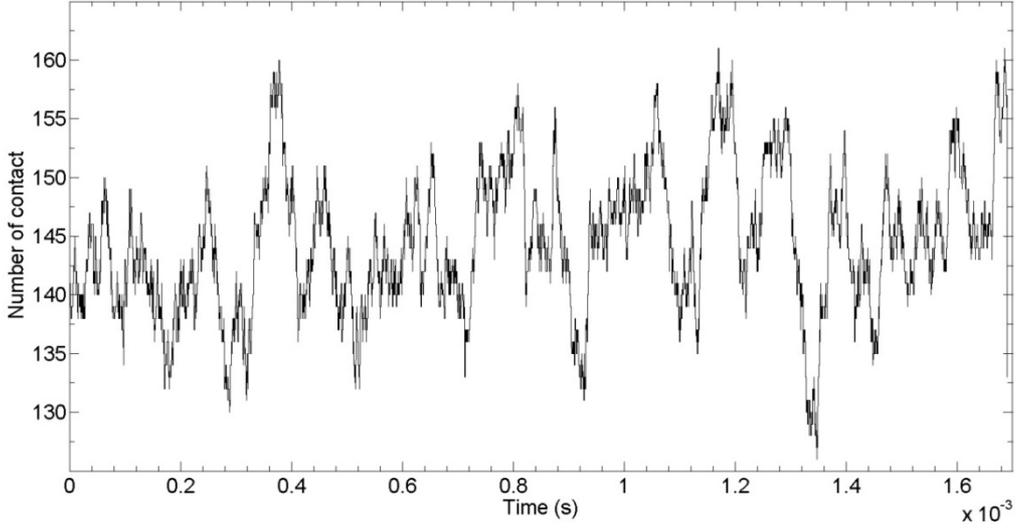

FIG. 3. Example of contact noise.

**III. The electrical model**

From the contact noise defined previously, we associate an electrical parameter to the contact spot. The sliding interface is now as a huge electrical parallel circuit wherein each contact spot is an independent electrical device – $Z_i$ – with a switch – $K_i$ – (fig. 4).

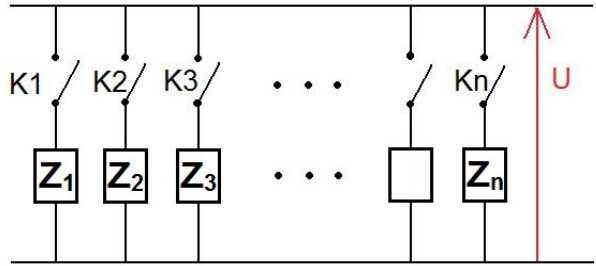

FIG. 4. Schematic visualization of the sliding interface from an electrical point of view. $K_1$, $K_2$, …, $K_n$ are the switch functions associated to each spot.

Thus, for a given number of contacts into the interface N(t) during sliding, we can associate an impedance value to the interface as follow:

$$\frac{1}{Z(t)} = \sum_{i=1}^{N(t)} \frac{1}{Z_i} \leftrightarrow Y(t) = \sum_{i=1}^{N(t)} Y_i \qquad \text{[Eq. 04]}$$

If we consider that $Y_i \approx \bar{Y_i}$, it yields:

$$Y(t) = N(t)\bar{Y_i} \qquad \text{[Eq. 05]}$$

At null speed the contact number remains constant and we consider $\overline{Y_t}$ constant during sliding. It yields:

$$Y_{(V=0)} = N_{(V=0)} \overline{Y_t} \qquad \text{[Eq. 06]}$$

From Eq. 05 and Eq. 06 we can deduce the relation between mechanical and electrical behavior of the interface as:

$$R(t) = \frac{N_0}{N(t)} R_0 \qquad \text{[Eq. 07]}$$

Finally, we link the electrical aspect to the mechanical one through the relationship of the resistance of one spot and the contact noise piloted by mechanical parameters. An example of the resistance noise from the sliding interface is shown on figure 5. Note we choose an arbitrary initial contact number $N_0$ to focalize on the behavior of interface resistance according to the functionality parameters. Note too that choose $N_0$, leading to a constant $\overline{Y_t}$ during sliding, is similar to choose a constant mean surface spot during sliding, independently from the influence of the sliding speed.

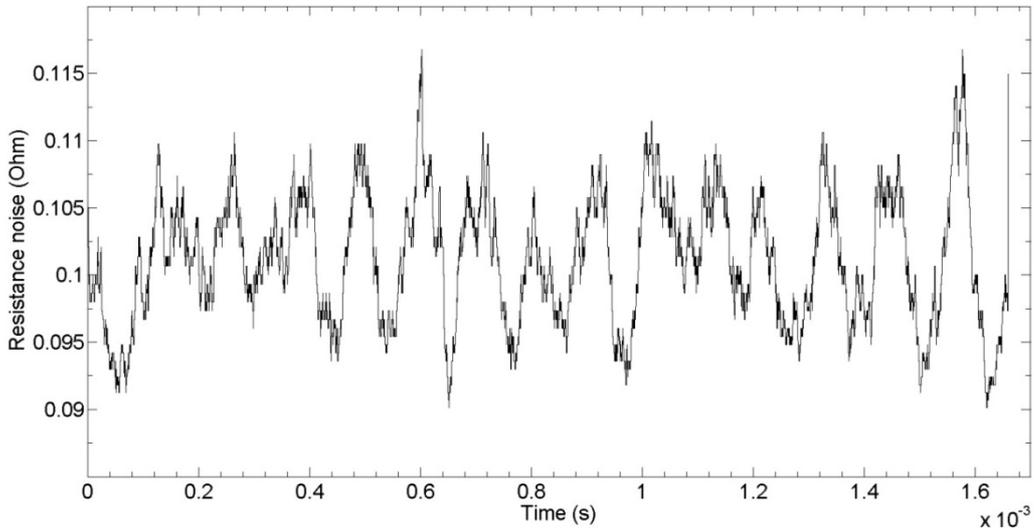

FIG.5. Example of resistance noise for a constant sliding speed and for an initial contact number equal to 200.

## IV. RESULTS, INTERPRETATION AND DISCUSSION

In this section are presented the numerical results of the simulation. In a first step we will focus on the interface resistance from the contact noise versus the sliding speed. Results will show for a specific choice of parameter a closed behavior from experimental data. In a second step we will focus on the spectral properties of the resistance noise and will also compare them to experimental data. In the last case we will make a strong link between the electrical measurements and the mechanical properties through the interpretation of the relation of the magnitude order of the power spectral density versus the ratio of contact number used in the interface.

Note that in the following results we use an arbitrary parameter that we call $N_0$ which must be understood as the contact number able to interact inside the sliding interface and are the total number of contacts at null speed ($\omega_r = 0$).

## A. Interface resistance

As we saw before, the mechanical contact behavior leads the electrical one: voltage noise is directly influenced by the contact noise. So, we compute the ratio between the mean contact number inside the interface during sliding and the $N_0$ parameter versus the sliding speed and show it on the left of the figure 6. Next, we study the voltage noise versus the sliding speed for different value of the $N_0$ parameter, showed on the right of the figure 6.

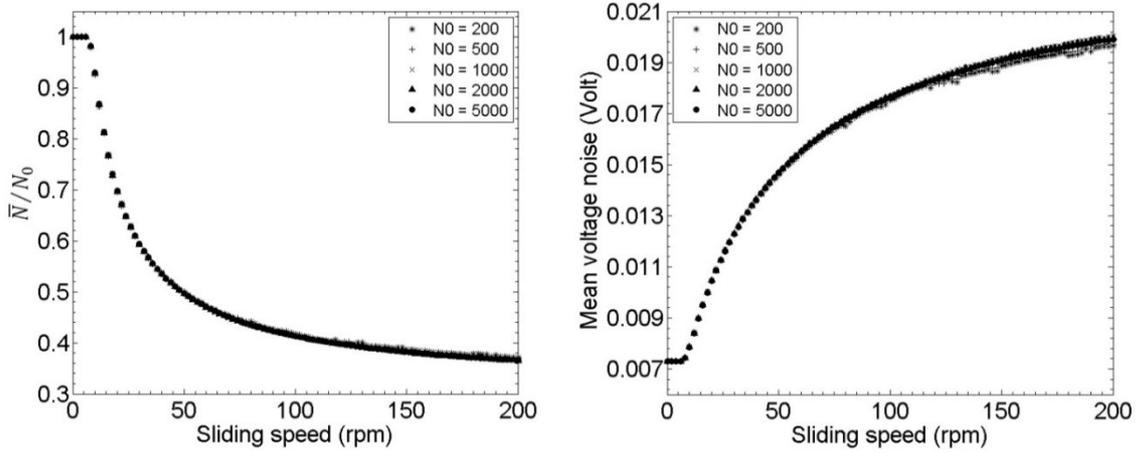

FIG. 6. Left: Ratio $\bar{N}/N_0$ vs. the sliding speed. Right: Mean voltage noise vs. the sliding speed (computed from the resistance noise with a fixed current equal to 100 mA).

As we can see, the ratio $\bar{N}/N_0$ decreases when the speed increases. The increasing of this ratio leads to an increasing of the interface resistance. As we suspect, sliding speed and steady force has an antagonist role: stronger the steady force, better the electrical contact, and so, faster the sliding speed, worse the electrical conductance. Note that this behavior of the ratio $\bar{N}/N_0$ is independent of the fixed parameter $N_0$.

The comparison between experimental and numerical results (fig. 7) shows that numerical results reach the experimental results magnitude order, and are close to them between 20 and 100 rpm. Divergence beyond 100 rpm seems to indicate that the mechanical model as it is built is not enough to describe the phenomenon for high speed: we must take into account that in the experimental measurements, we cannot control the modification undergone by the sliding interfaces like wear [15]. In [5] Chevallier et al. show that the voltage noise grows with aging, so for high sliding speed the evolution of the surface state is probably accelerated. Each turn of the ring modified the surface state, so the contacts of one surface never meet the same contacts again during the time: in the model we fixed the roughness of our profile and it remains constant, which is not a quite realistic behavior. Beyond 100 rpm, thermic effects can probably appear, leading to an increasing of the resistance interface.

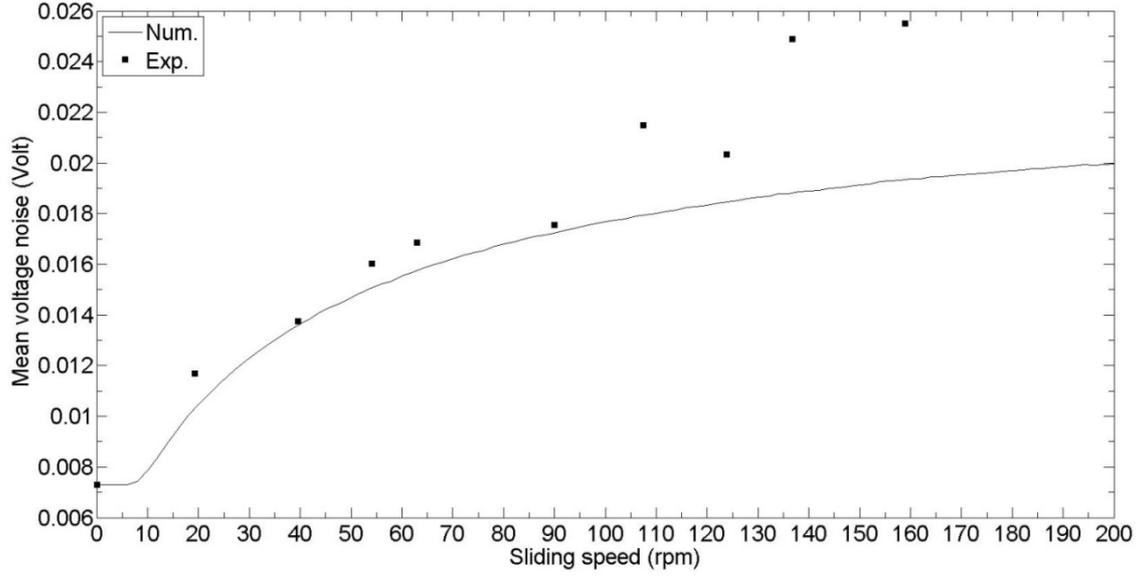

FIG. 7. Numerical and experimental results of the voltage noise versus the sliding speed (profile roughness fixed around 20μm).

## B. P.S.D. Properties

We compute the magnitude order of the P.S.D. from the voltage noise versus the ratio $\bar{N}/N_0$ (fig. 8) from our numerical study. We clearly see a specific behavior led by the mean contact number inside the contact interface.

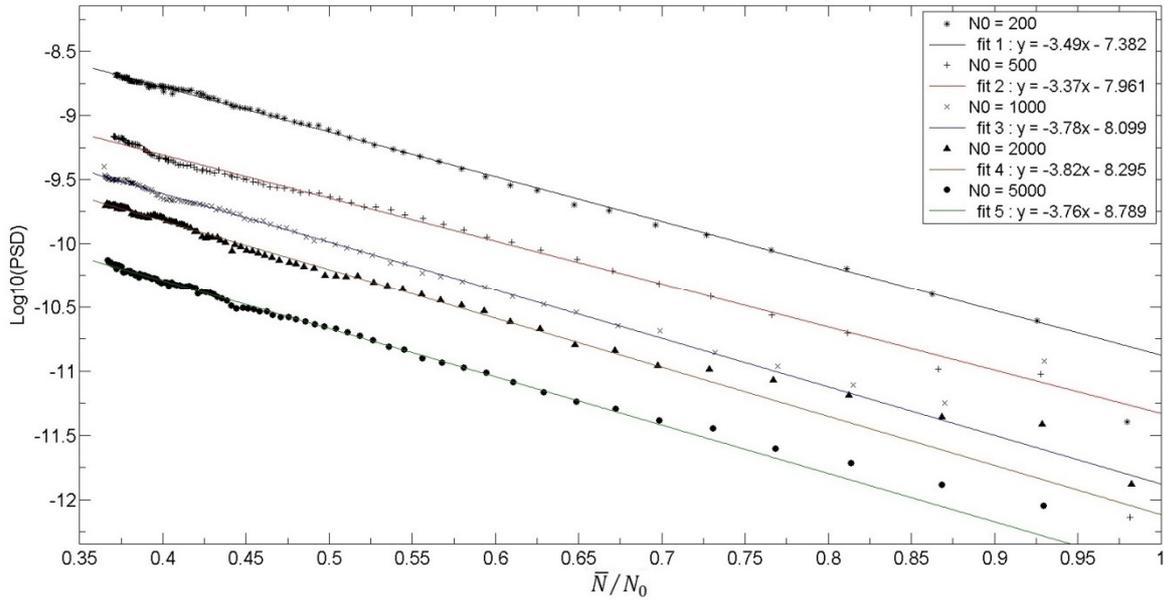

FIG. 8. Magnitude order of the P.S.D. according to ratio $\bar{N}/N_0$.

First, for each line, the slope seems to be the same: on average, the value of the slope is around 3.644 (+/- 7.6%). It's meaning a specific behavior, characterized by a parameter which is independent from the parameter $N_0$. Next, the intercept is determined by the parameter $N_0$ and is inversely proportional to this one. From the figure 6 we deduced a saturation of the ratio $\bar{N}/N_0$ in high speed (above 0.35 for 200 rpm), so our intercept can be interpreted as the maximum value of the $\log_{10}$(P.S.D.) for an infinite sliding speed.

From these results, we can expressed a behavior law: the P.S.D. of the voltage noise is leading by a same parameter independent from the initial condition – the slope –, and by another parameter which is piloted by the $N_0$ parameter – the intercept –.

It yields:

$$P.S.D. = A_{(N_0)} \cdot 10^{(\alpha \bar{N}/N_0)} \qquad \text{[Eq. 08]}$$

Where A is a function of $N_0$, and α the slope equal to 3.644.

We compare now our numerical results to the experimental ones, and get the figure 9. The number $N_0$ inside the experimental sliding interface is around 2000: experimental as numerical values are much closed. We see also that below a certain value of the ratio $\bar{N}/N_0$ (around 0.38), i.e. beyond a certain value of sliding speed (around 150 rpm), the experimental values jump on of almost a magnitude order. We fitted the experimental results for main values (fit 2) and for the firsts values (fit 3), and compared them to the fitted numerical values (fit 1). The value of the slopes of the fit 1 and the fit 2 is almost identical, there is just a difference between them about 18%. The value of the slope of fit 3 is less important about 33% from the fit 1 slope value. From the fit 3 we can deduce the parameter $N_0$ associated to it: we estimate in this case $N_0$ is equal about 350 contacts.

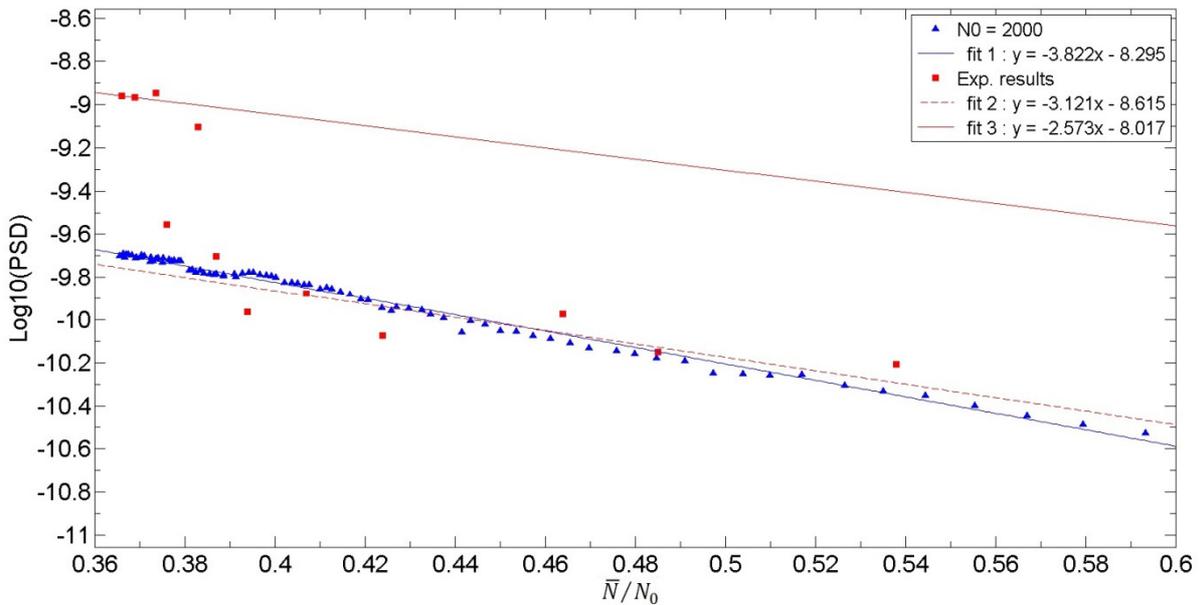

FIG. 9. Comparison between numerical and experimental results of the P.S.D. magnitude order, according to the ratio $\bar{N}/N_0$.

In accordance with the literature [14, 16], the real contact number between two surfaces is determined by the steady force and surface roughness. From the Holm model [1] we can estimate the mean surface radius of one of these 2000 contacts about few Angstroms. It's clear that $N_0$ cannot be understood as a reliable contacts number. These last results show us clearly a significant influence of the interaction between the sliding surface spots on the spectral properties of the voltage noise. Beyond 150 rpm, the experimental values seem to indicate a new process which must be understood. A first hypothesis is to consider the influence of the scale of the surface fluctuation as a functionality limit for a specific range of sliding speed.

**V. CONCLUSION AND PERSPECTIVES**

Our rudimentary sliding contact modeling allowed us to establish a first series of numerical results comparable to the experimental ones. These results indicate that in the interface mean voltage case, the magnitude order of the values is much closed of the experimental one in a given sliding speed range. So, this first study characterizes the sliding interface through the electrical resistance that it induces by the mechanical interactions of its asperities, making a direct link between mechanical and electrical properties of the surface. We saw also that the contact number inside the interface is important because it has a significant influence on the spectral properties: the behavior of the voltage noise P.S.D. is directly piloted by the contact number between the sliding surfaces, as shows the figure 9. Note that if here $N_0$ is used arbitrarily, it opens perspectives to study the influence of the voltage noise versus the steady force, because contact number is determined by the surface roughness and the force between the sliding surfaces [1, 10, 13, 14].

In the case of the P.S.D., the comparison between experimental and numerical results allowed us to distinguish a specific behavior: a "jumping" effect for high sliding speed range. Thus, we characterize the spectral properties of the voltage noise, according to the mechanical parameters through the computation of its P.S.D. This study highlights an important parameter, $N_0$, which remains to define more exactly: the presence of a shift of values beyond a certain sliding speed adds a complexity related to the phenomenon interpretation.

If our model seems to be agree with the experimental measurements, we must keep in mind that our approach neglects some other influences: others phenomena at the interface contribute to the voltage noise, like wear production, or the time evolution of roughness. If all the phenomena at the interface cannot be take account at this time, one important parameter must be added in our following studies: temperature. Knowing pertinently that temperature influences the electrical resistance, it is necessary to add this parameter to establish a more pertinent comparison to the experimental results.

**ACKNOWLEDGMENT**

This study has been carried out thanks to the financial support provided by the Region of Picardie.